\documentclass[usenatbib,onecolumn,usegraphicx]{mn2e}

\title[Micro-images of lensed quasars]
{Understanding micro-image configurations\\ in quasar microlensing}

\author[P. Saha \& L.L.R. Williams]
{Prasenjit Saha$^1$\thanks{psaha@physik.uzh.ch} and 
Liliya L.R. Williams$^2$\thanks{llrw@astro.umn.edu}\\
$^1$Institute for Theoretical Physics, University of Z\"urich, 8057
Z\"urich, Switzerland\\
$^2$School of Physics and Astronomy, University of Minnesota, 116
Church Street SE, Minneapolis, MN 55455 USA}

\begin{document}



\maketitle

\def\bkappa{\bar\kappa}
\def\bgamma{\bar\gamma}

\label{firstpage}

\begin{abstract}
  The micro-arcsecond scale structure of the seemingly point-like
  images in lensed quasars, though unobservable, is nevertheless much
  studied theoretically, because it affects the observable (or macro)
  brightness, and through that provides clues to substructure in both
  source and lens.  A curious feature is that, while an observable
  macro-image is made up of a very large number of micro-images, the
  macro flux is dominated by a few micro-images.  Micro minima play a
  key role, and the well-known broad distribution of macro
  magnification can be decomposed into narrower distributions with
  $0,1,2,3\ldots$ micro minima.  This paper shows how the dominant
  micro-images exist alongside the others, using the ideas of Fermat's
  principle and arrival-time surfaces, alongside simulations.
\end{abstract}

\begin{keywords}
gravitational lensing: micro
\end{keywords}

\section{Introduction}

In multiply-imaged lensed quasars, each observable ``macro'' image is
composed of many unresolved ``micro'' images, due to stellar-scale
substructure in the lensing galaxy or cluster.  The micro-images can
vary in response to even the tiny proper motions of stars or galaxies at
cosmological distances, causing occasional rapid changes in the
observed macro flux.

Much theoretical and numerical work has been done on quasar
microlensing, since the original prediction of
\cite{1979Natur.282..561C}.  The most common strategy, introduced
originally by \cite{1981ApJ...244..756Y} is ray-tracing: one sets up a
random star field with mean density and external shear according to
some macro model of the lens, and traces rays backwards to the source
plane to compute a macro magnification pattern. Recent examples include
random motions of stars, with a view of estimating the properties of the
microlensing stars and the accretion disk of the quasar source
\cite[e.g.,][]{2010ApJ...712..658P,2010ApJ...712..668P}. 
Micro-images do not appear explicitly in this technique.
A contrasting approach is to find all the micro-images due to a random
star field and add up their fluxes.  \cite{1986ApJ...301..503P} was
the first to do so, and emphasized the large number of micro-images.
Subsequent work produced some elegant image-finding algorithms
\citep{1993ApJ...403..530W,1993MNRAS.261..647L} and further insight
into the nature of micro-images. A recent evaluation of microlensing 
computational techniques can be found in \cite{2010arXiv1005.5198B}.

Yet the macro flux appears to be dominated by a few images.
\cite{1984JApA....5..235N} argued, theoretically from the statistical
properties of a random star field in an external lensing shear, that
the formation and merging of a single pair of macro-images would cause
a sharp change in the macro flux.  Numerical studies of simulated star
fields by \cite{1992ApJ...386...30R} verified this
behaviour. \cite{2002ApJ...580..685S} considered this further,
emphasizing that macro minima and saddles behave differently and
supplying a toy model for understanding the difference.
\cite{2003ApJ...583..575G} showed that the number of micro minima is
well approximated by a Poisson distribution with mean proportional to
the macro magnification.

Such considerations lead to the questions: how do the dominant micro
images co-exist alongside the many more numerous faint micro-images?
And what are the consequences for the brightness of the macro-images?
In this paper we attempt to answer these questions, by considering an
aspect that has previously not received much attention in the context 
of quasar microlensing: the geometry of the arrival-time surface.

\section{Image configurations}

The usefulness of the arrival-time surface is that images, be they of
the macro or micro variety, form at its local extrema: minima, maxima
(positive-parity images), and saddle points (negative parity).  Of
all the arrival time contours, the skeletal ones are those that are
self-intersecting and hence correspond to saddle points
\citep{1986ApJ...310..568B}.  Positive parity images form inside loops
created by self-intersecting contours.  Saddle-point contours are
helpful in understanding the macro-image configuration of lensed
quasars \citep[e.g.,][]{2003AJ....125.2769S} and have also been useful
for studying unusual macro systems \citep[see figures in][]
{2001ApJ...557..594R,2003ApJ...590...39K}.  As we will show, they are
also useful for understanding micro-image configurations.

When the matter distribution is continuous, a macrolens generates
macro-images whose macro magnification is
\begin{equation}
   \bar M^{-1} = (1-\bkappa)^2 - \bgamma^2,
\end{equation}
where $\bkappa$ and $\bgamma$ are the local value of the convergence
and shear respectively.
The arrival time has the form
\begin{equation}
   \textstyle
   \tau(x,y) = \frac12 (1-\bkappa-\bgamma) x^2
             + \frac12 (1-\bkappa+\bgamma) y^2
\label{macroarriv}
\end{equation}
Now suppose the continuous matter is broken
up into many point-like lenses, keeping the macro convergence of this
random star field the same as before, $\bkappa$.  In other words, the
Einstein rings of individual stars cover $\bkappa$ of the area. The
constant external shear $\bgamma$, is the same as before, and the macro
magnification has not changed, however, at a random point in the star
field, there is no $\kappa$ but there is a varying $\gamma$ from the
combined effect of the star field and the external shear.  Hence
locally, the micro magnification will be
\begin{equation}
   M^{-1} = 1 - \gamma^2
\label{micromag}
\end{equation}
The arrival time now takes the form
\begin{equation}
   \textstyle
   \tau(x,y) = \frac12 (1-\bgamma) x^2 + \frac12 (1+\bgamma) y^2
             - \sum_i \theta_i^2 \ln \left| {\bf x}-{\bf x}_i \right|
\label{microarriv}
\end{equation}
where $\theta_i$ is the Einstein radius of the $i$-th star. The
sum over logarithmic terms indicates that the arrival time surface
has acquired a new local maximum at each star, though these will have
zero magnification. Many new saddle-point
contours also appear, leading to additional local saddle points, as
well as occasional minima.  All these are new micro-images. At first
glance, the arrival time surfaces can look rather complicated, but we
will now show that they can be interpreted by decomposing them into simple
building blocks, disregarding unimportant components and highlighting
the important ones, i.e., those that contain the brighter images.

\begin{figure}
\includegraphics[width=.35\hsize]{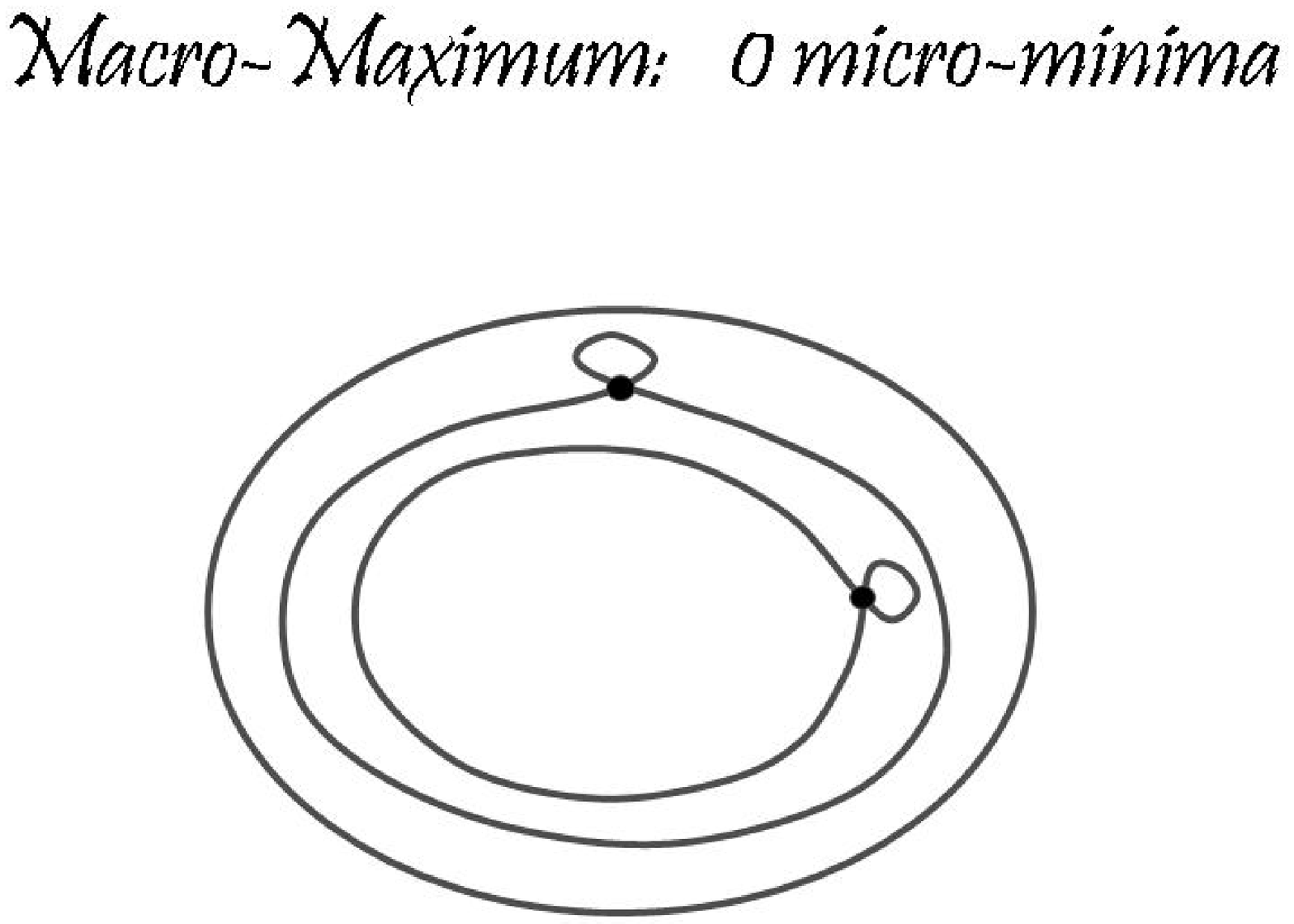}
\includegraphics[width=.35\hsize]{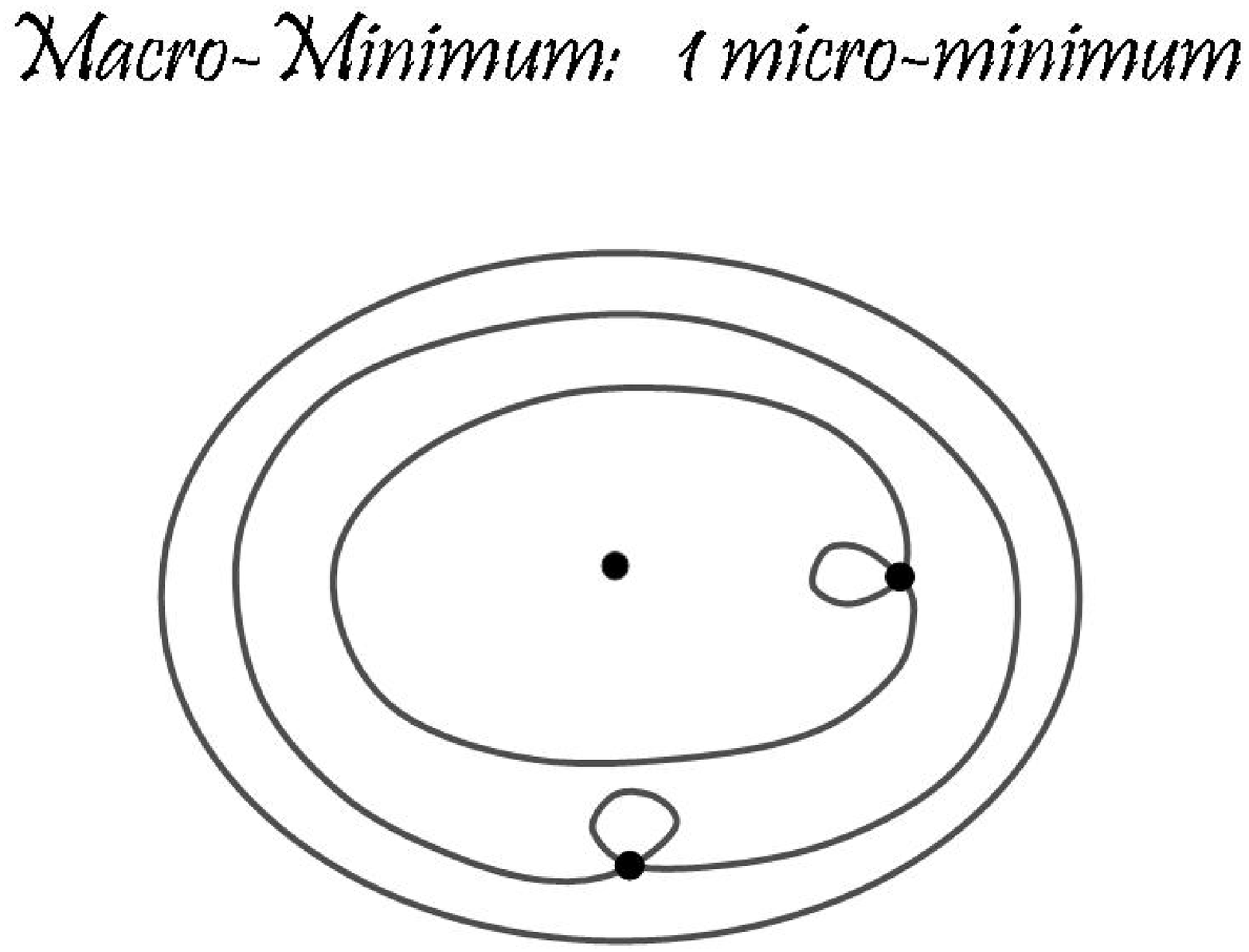}\\
\vskip5pt
\includegraphics[width=.35\hsize]{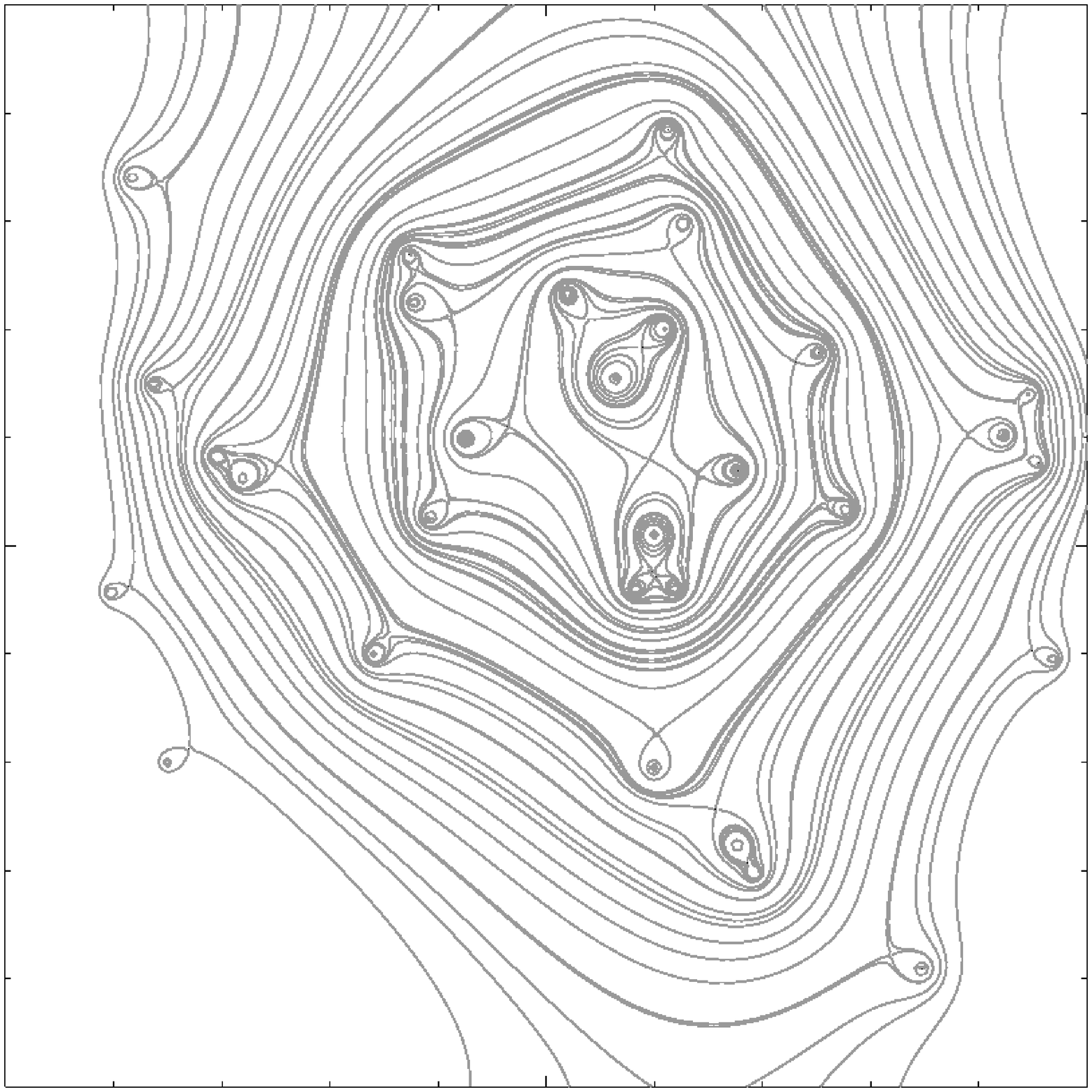}
\includegraphics[width=.35\hsize]{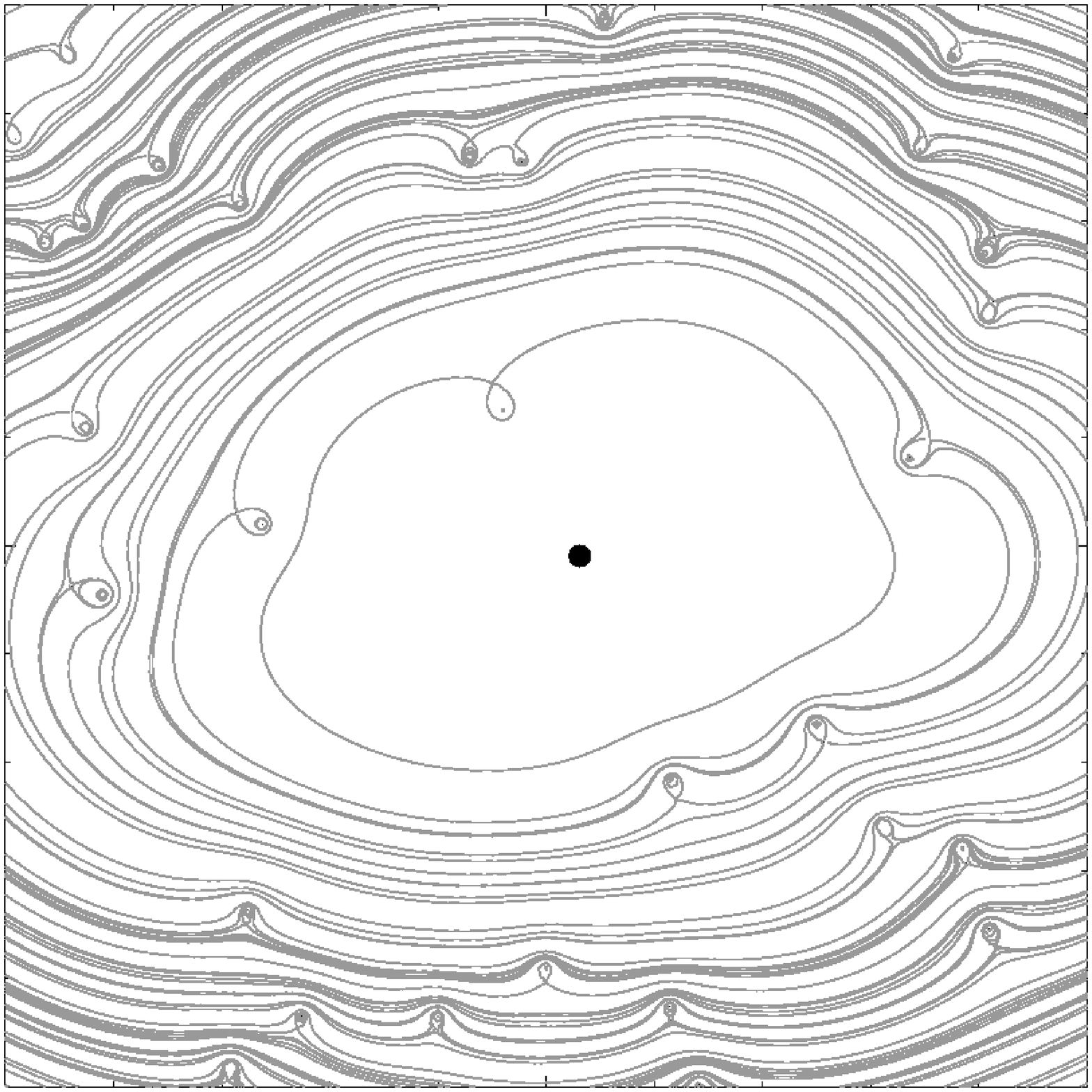}\\
\vskip10pt
\includegraphics[width=.35\hsize]{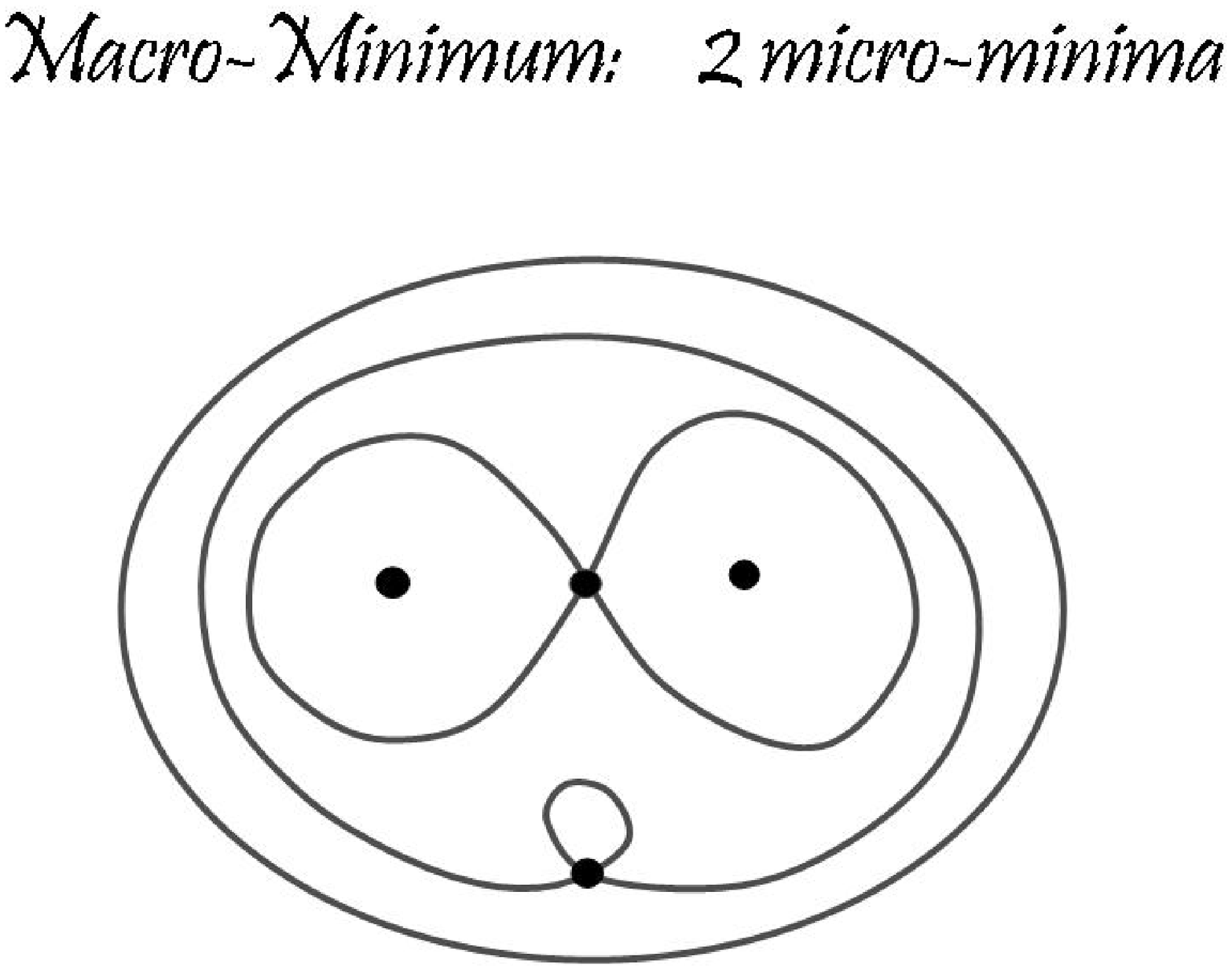}
\includegraphics[width=.35\hsize]{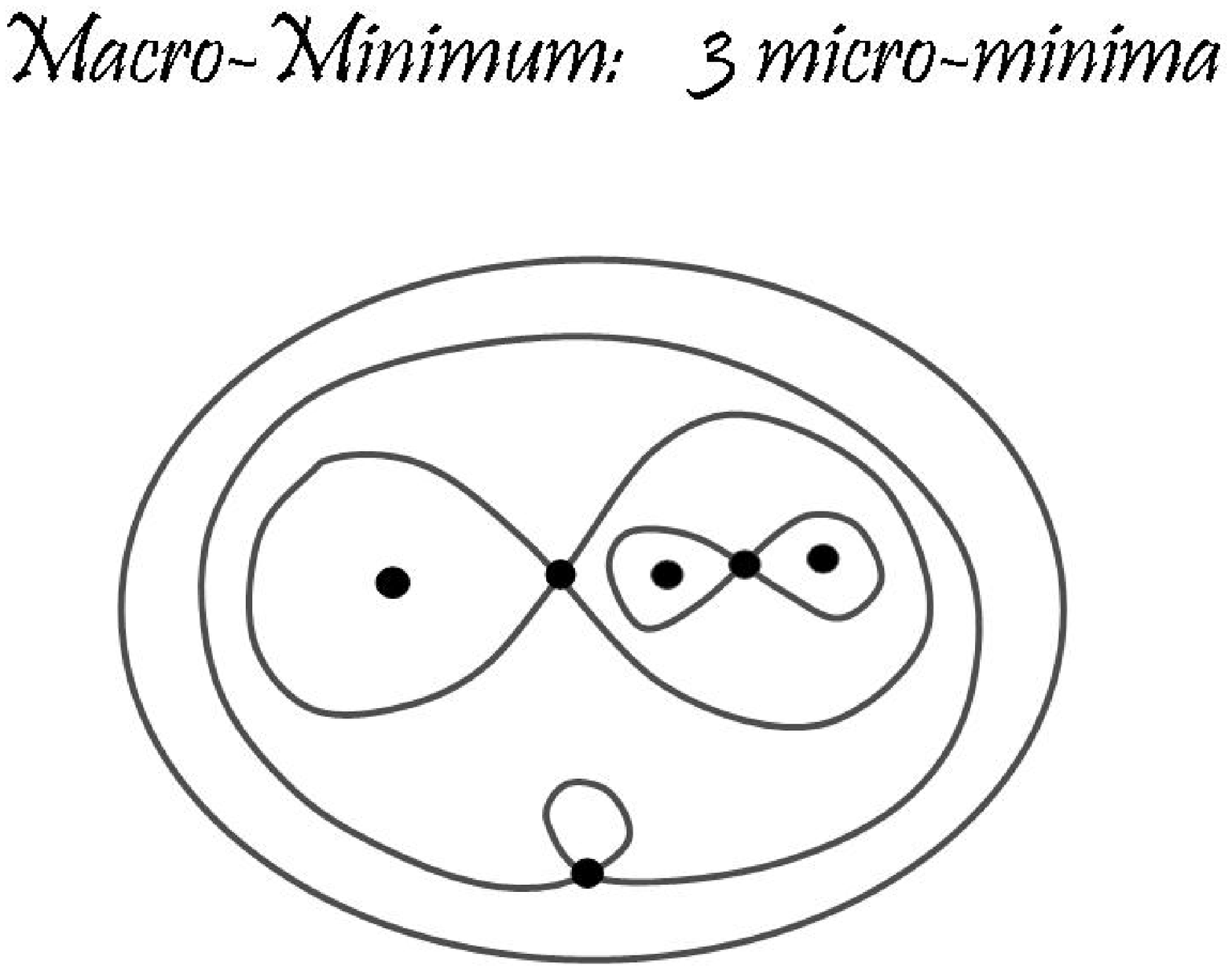}\\
\vskip5pt
\includegraphics[width=.35\hsize]{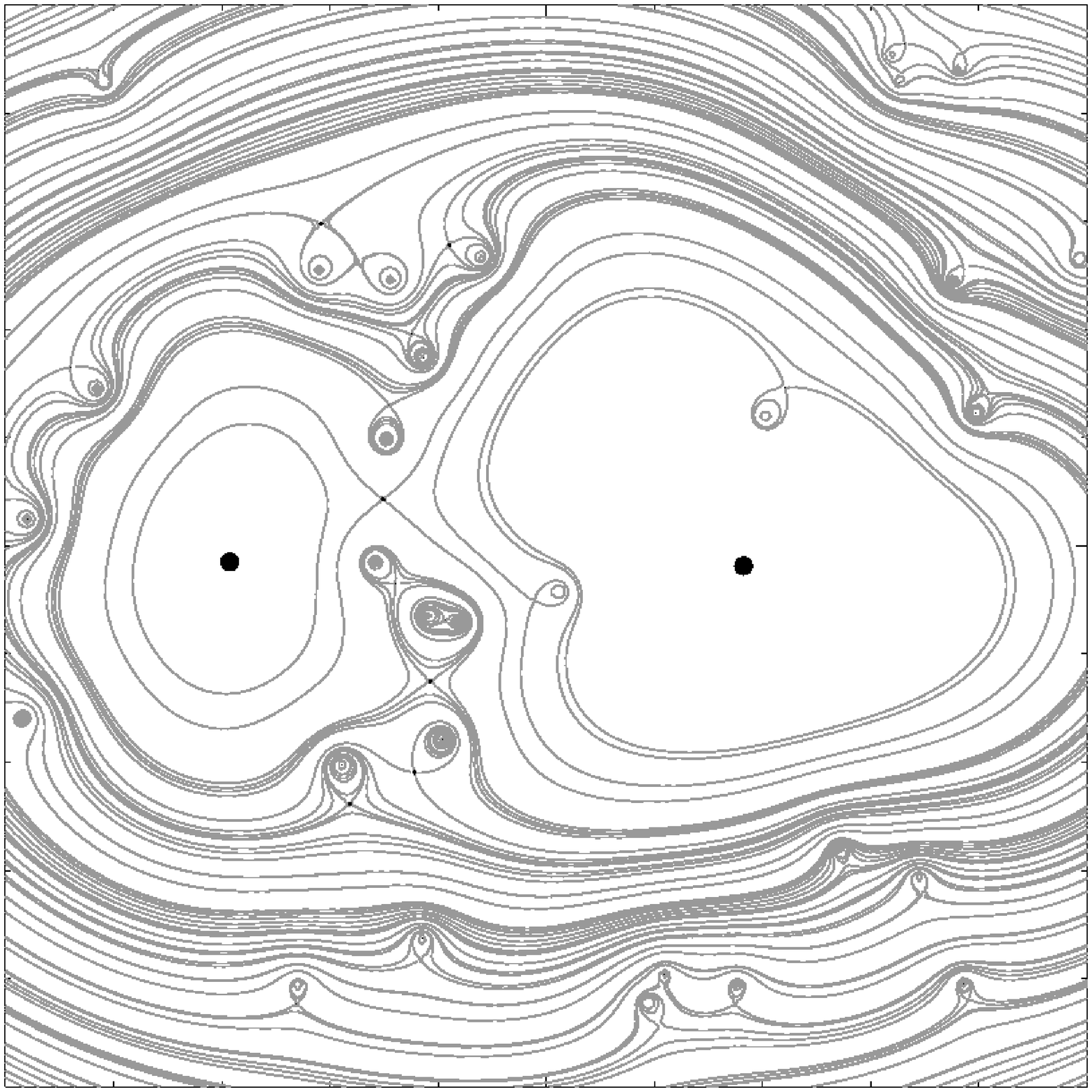}
\includegraphics[width=.35\hsize]{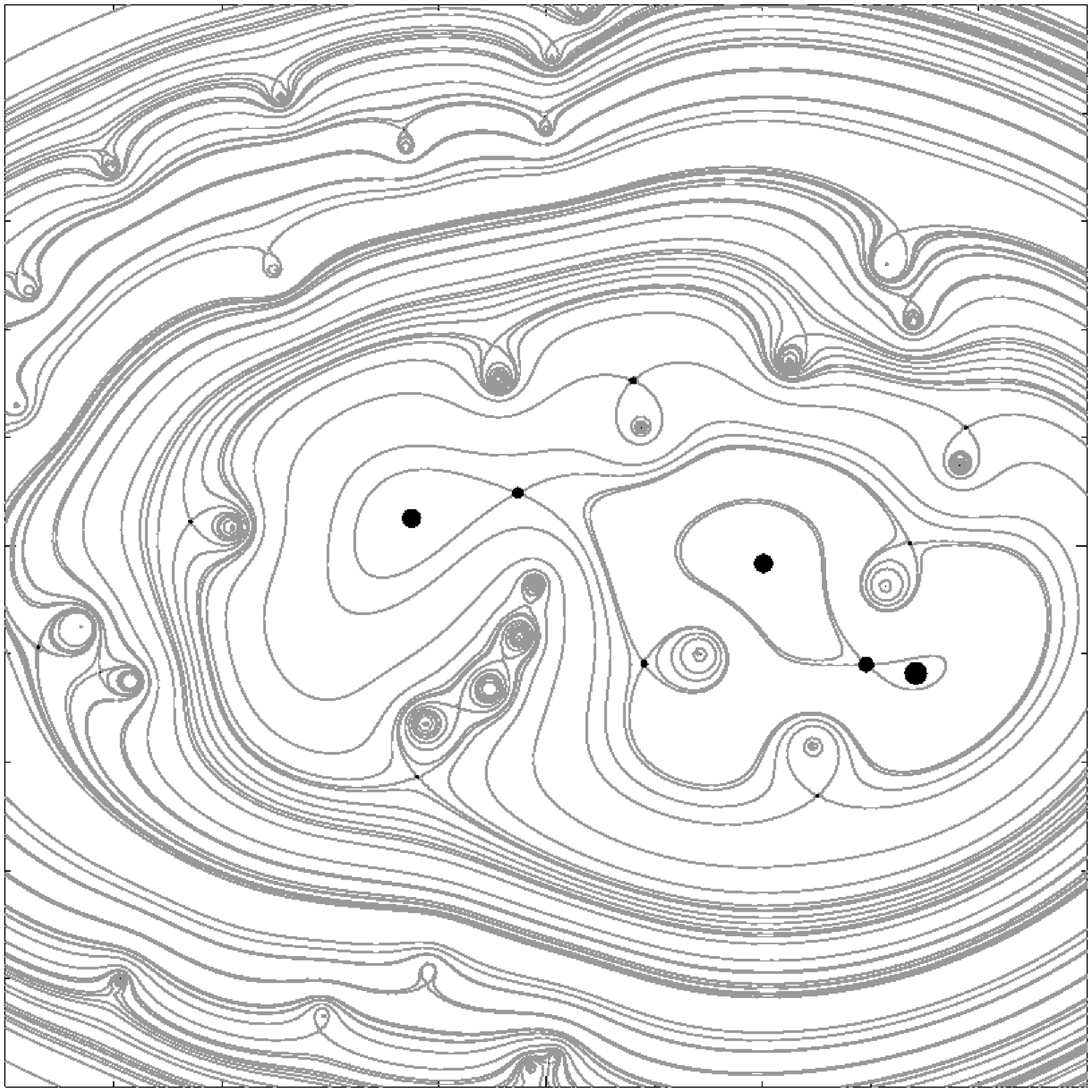}\\
\caption{Saddle-point contours for even macro parity.  Shown here are
  examples from simulations showing 0,1,2 and 3 micro minima.  The
  accompanying sketches give the qualitative features of the
  arrival-time surface.  Filled circles denote micro-images with
  significant flux. The area of these circles is proportional to the
  flux.}
\label{even-fig}
\end{figure}

\begin{figure}
\includegraphics[width=.35\hsize]{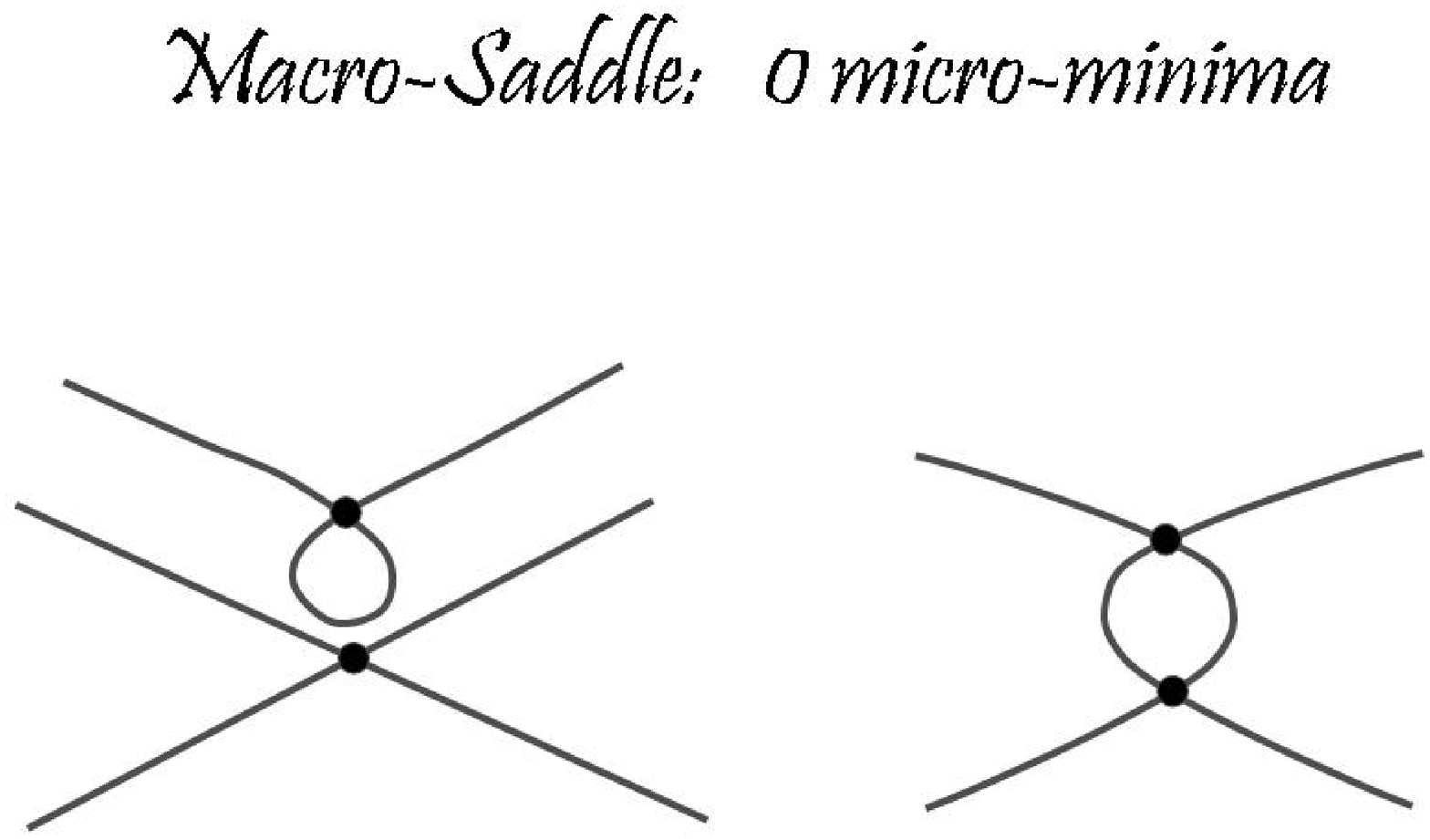}
\includegraphics[width=.35\hsize]{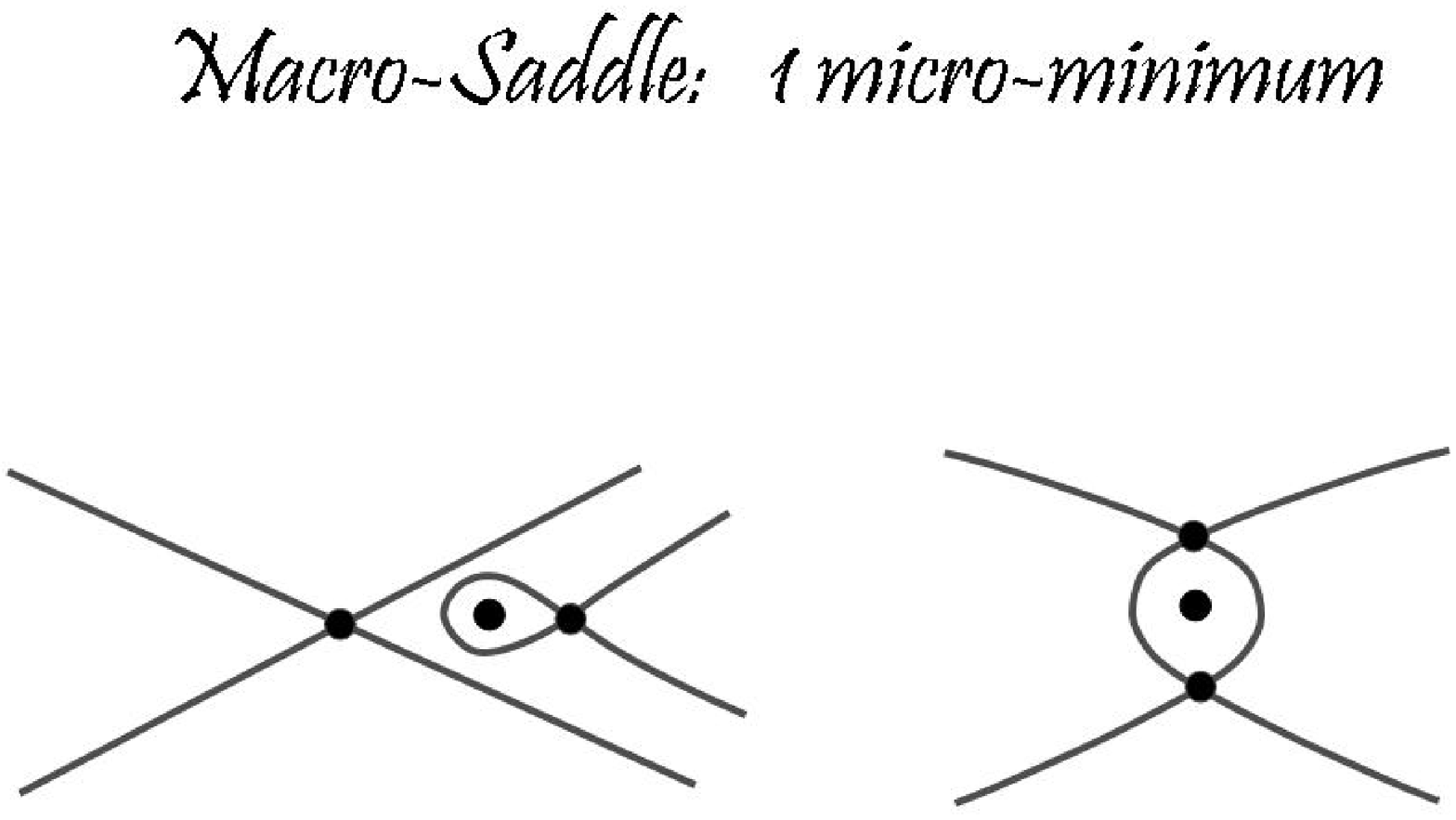}\\
\vskip5pt
\includegraphics[width=.35\hsize]{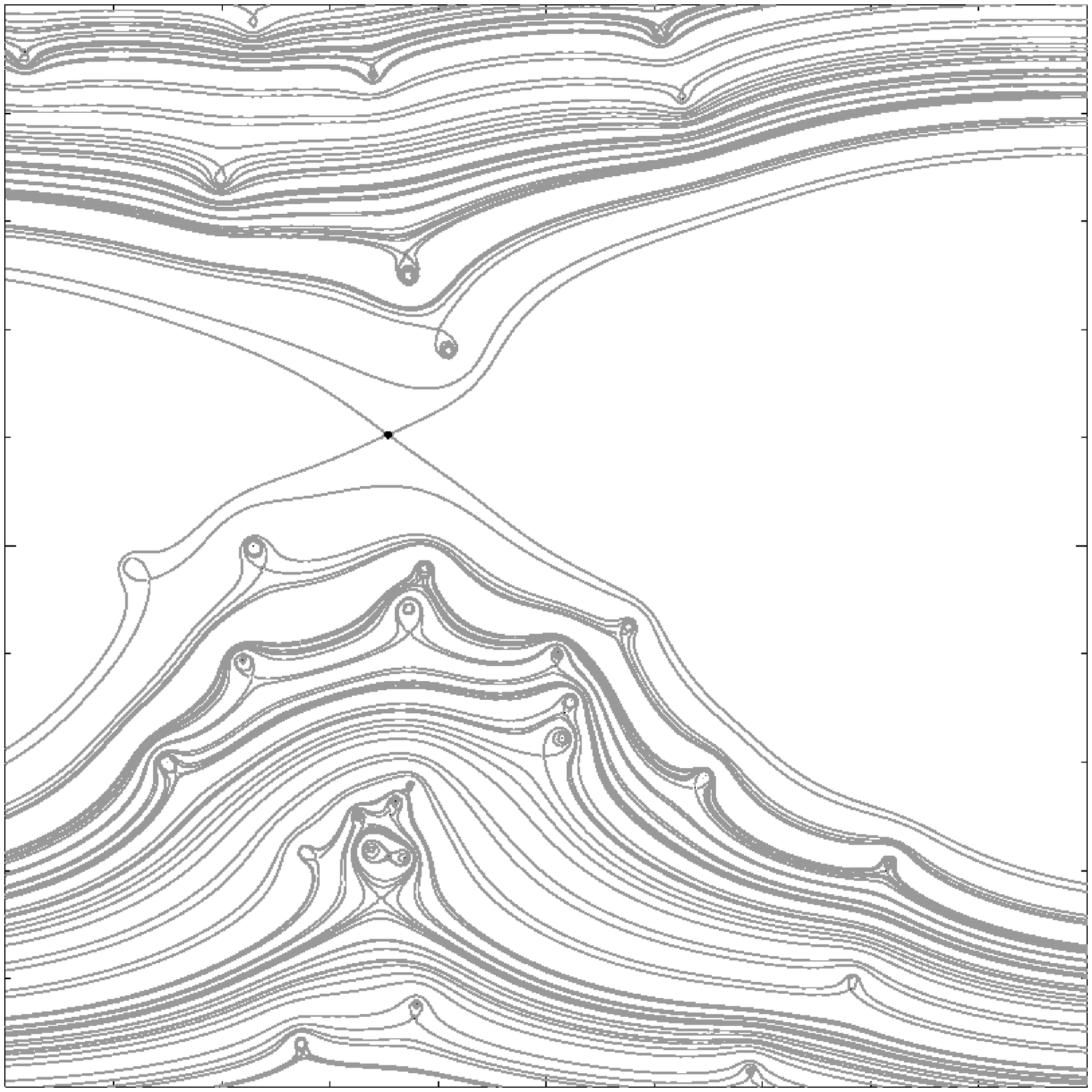}
\includegraphics[width=.35\hsize]{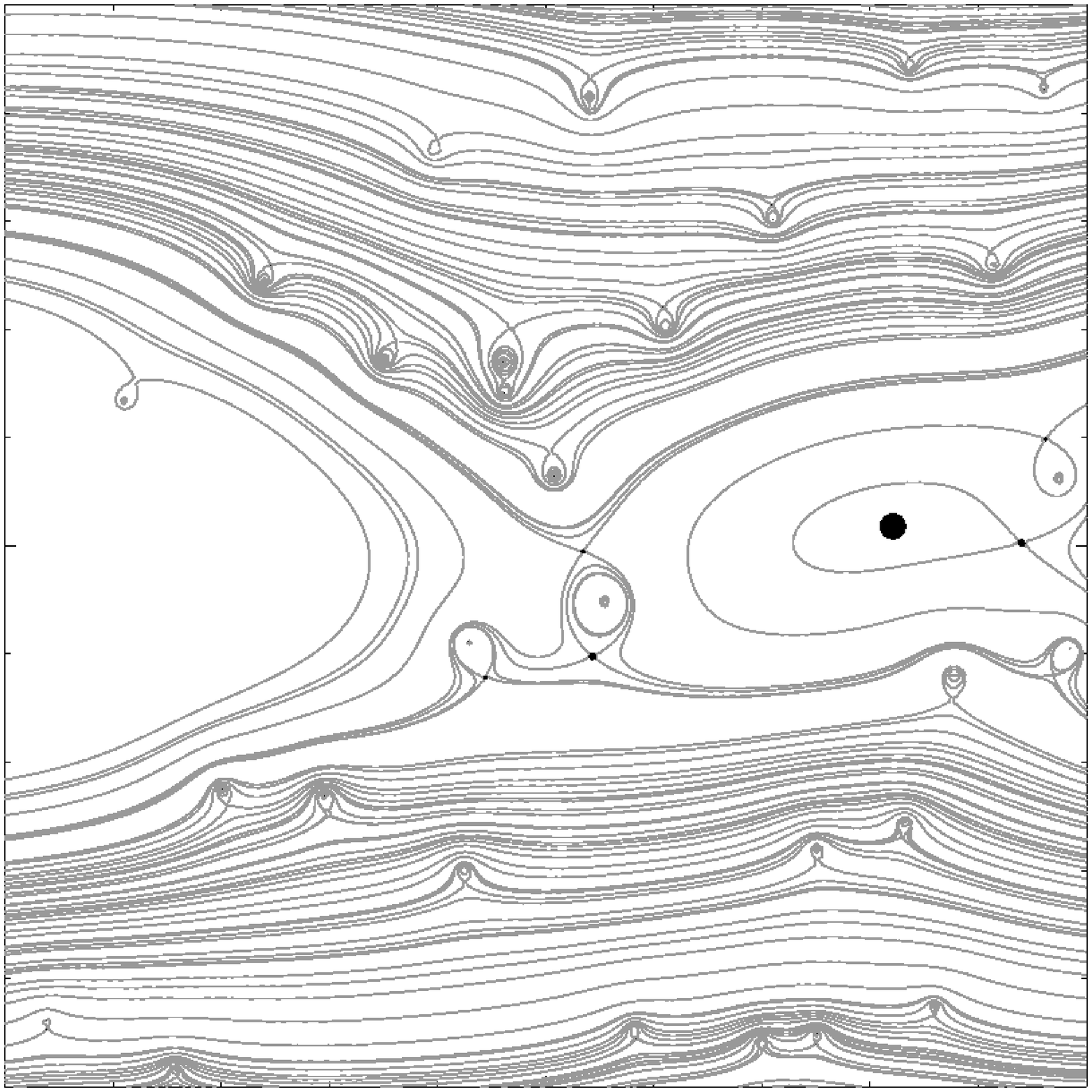}\\
\vskip10pt
\includegraphics[width=.35\hsize]{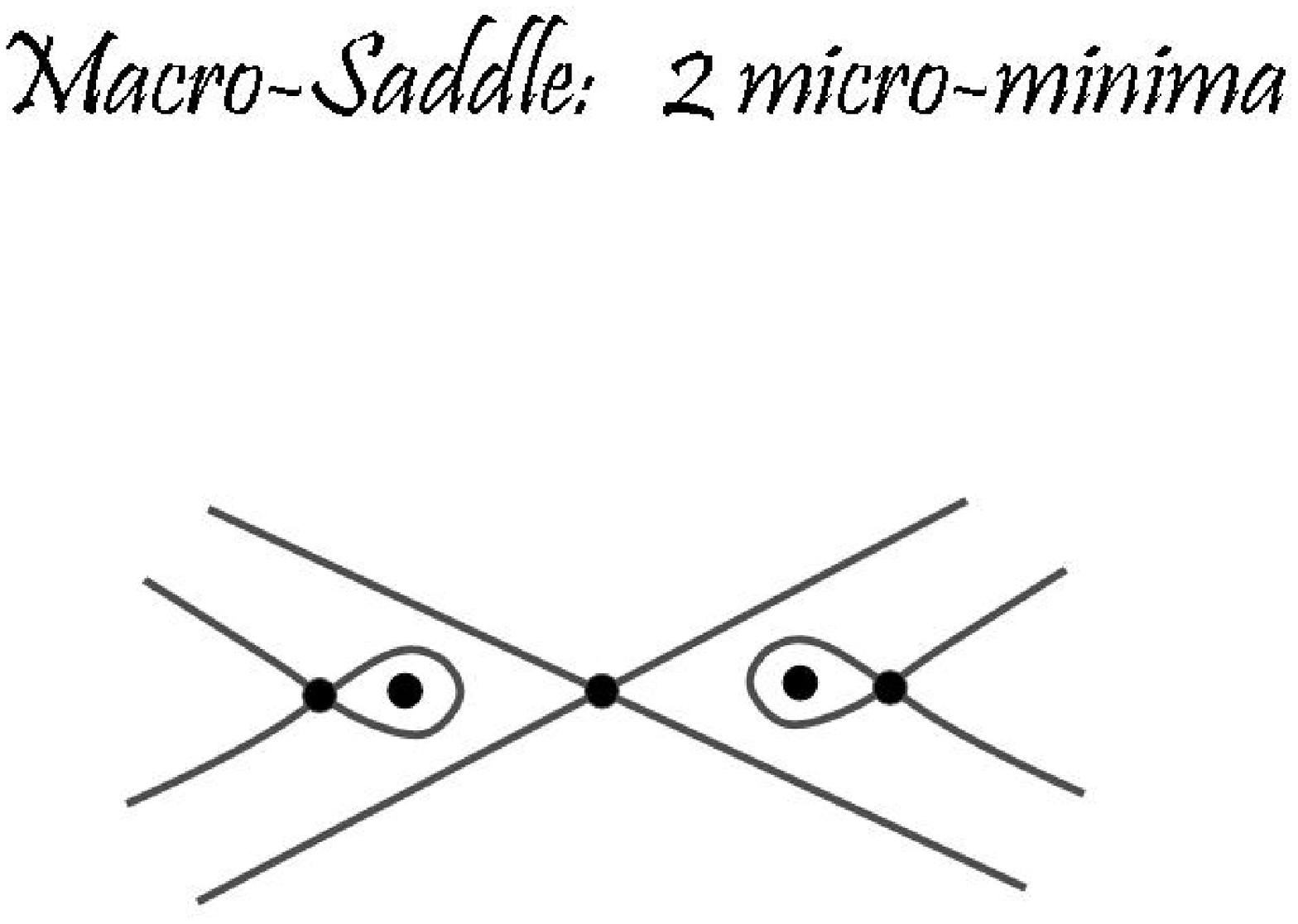}
\includegraphics[width=.35\hsize]{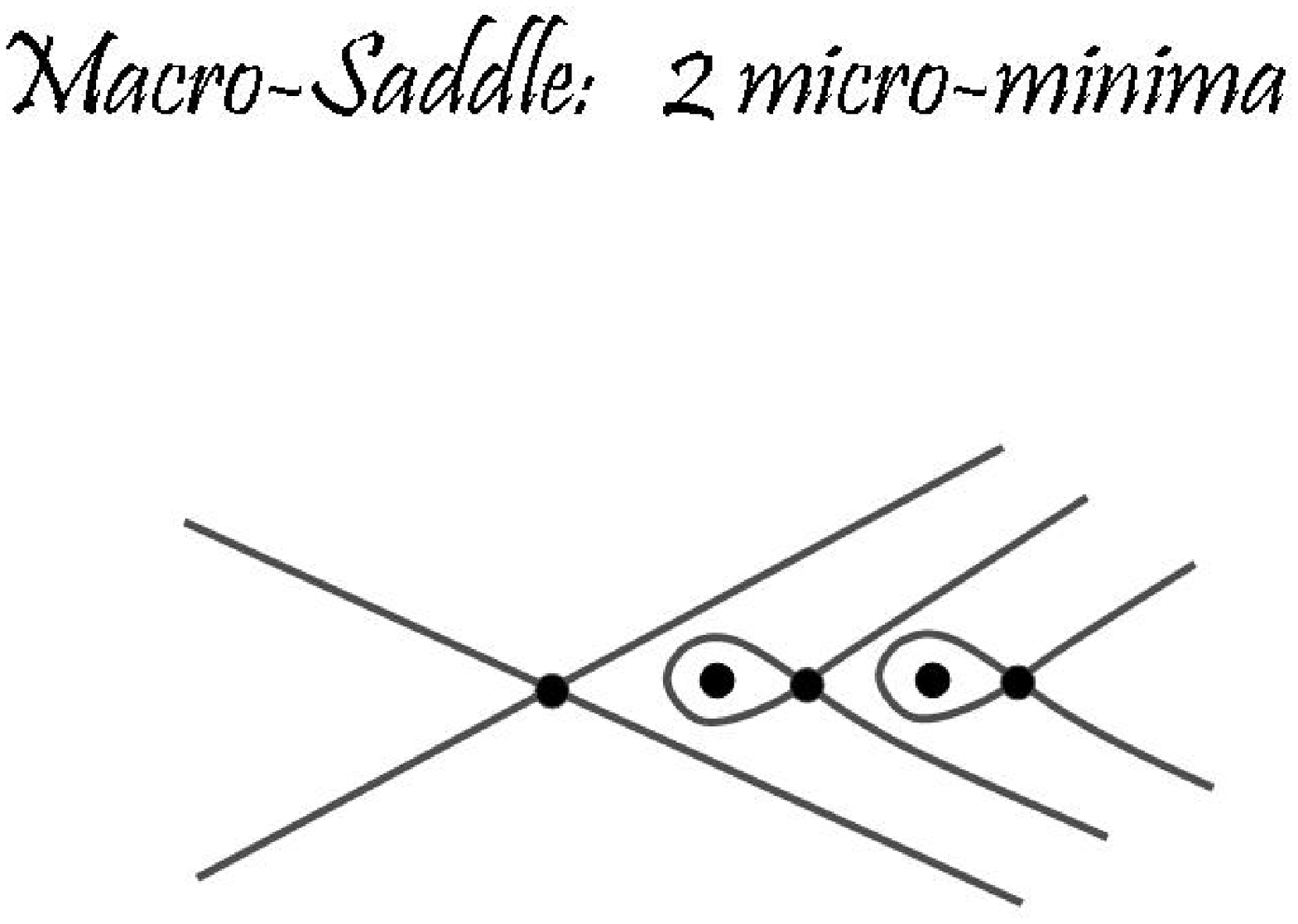}\\
\vskip5pt
\includegraphics[width=.35\hsize]{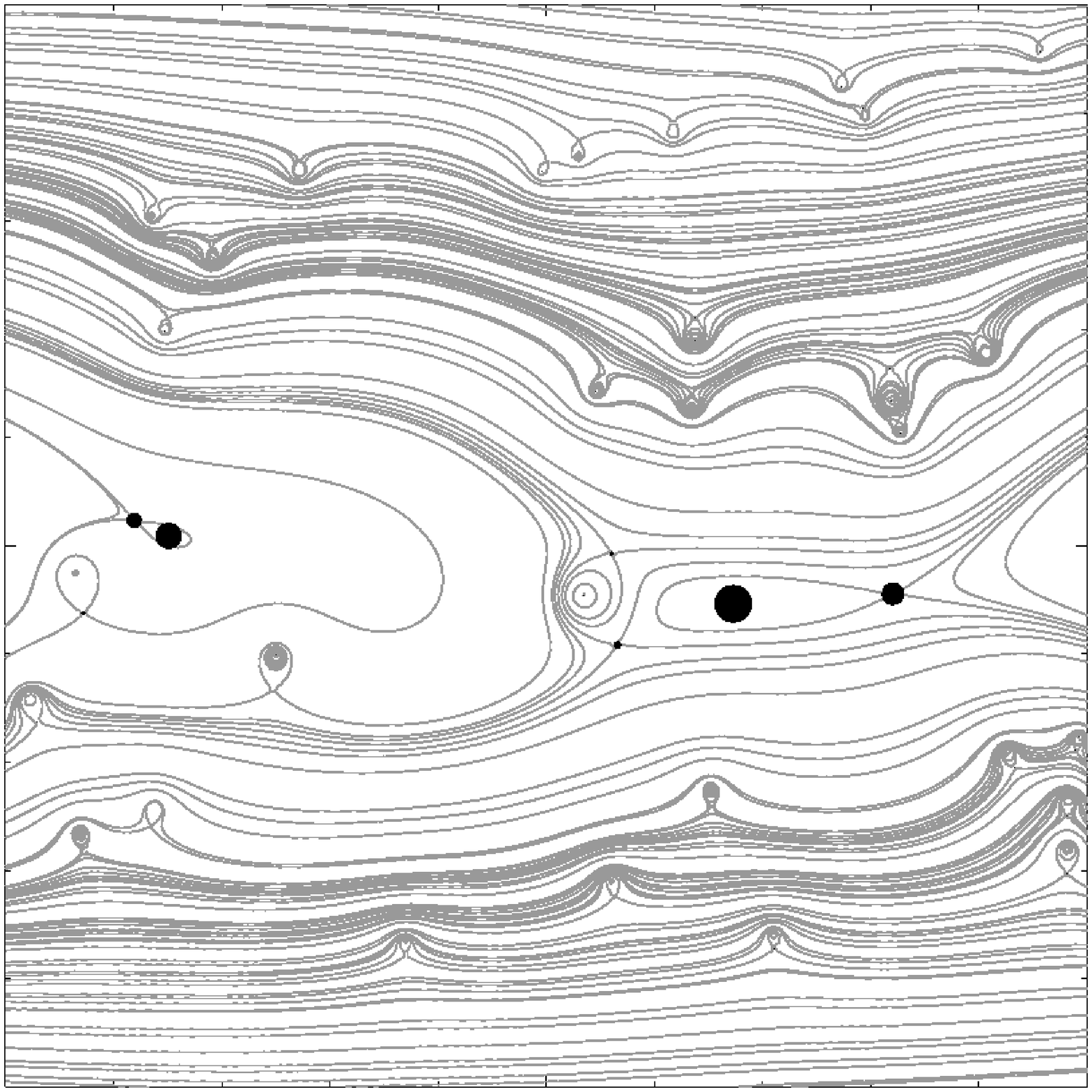}
\includegraphics[width=.35\hsize]{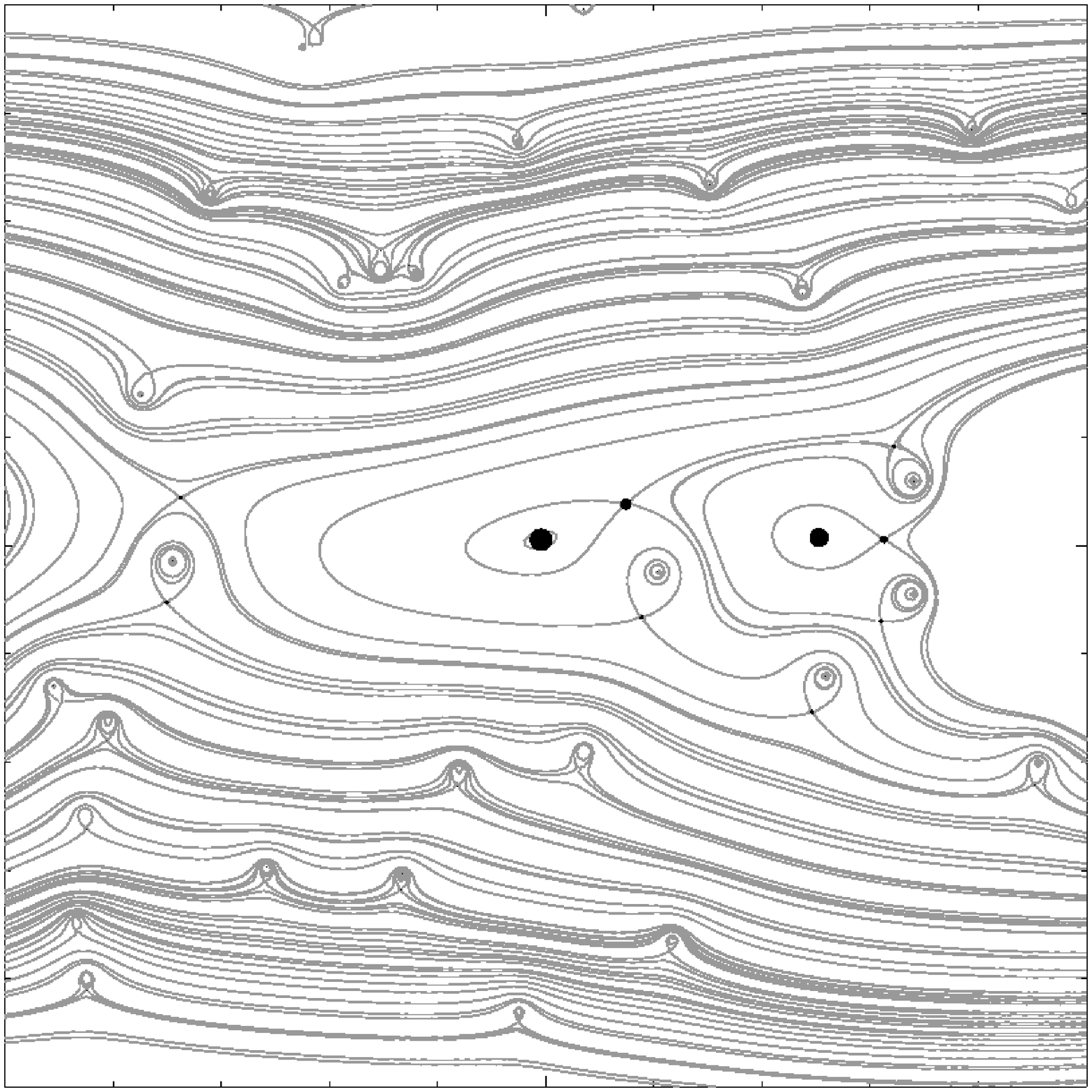}\\
\caption{Saddle-point contours for odd macro parity.  Shown here are
  examples from simulations showing 0,1, and 2 micro minima.  As in
  Figure~\ref{even-fig}, the accompanying sketches give the
  qualitative features of the arrival-time surface, while filled
  circles denote micro-images with significant flux, the area of these
  being proportional to the flux.}
\label{odd-fig}
\end{figure}

The decomposition of arrival-time surfaces is best explained
graphically.  This is done in Figures~\ref{even-fig} and
\ref{odd-fig}, which illustrate cases of even and odd parity respectively.

In Figure~\ref{even-fig} the top left panel displays a macro maximum.
The ``tree-rings'' plot in that panel is centred on the macro maximum,
and contains 30 randomly placed star lenses.  Micro-images were found
by using a recursive grid search. The curves are the saddle-point
contours, that is, those contours of the arrival time
(\ref{microarriv}) which pass through a saddle-point.  The
micro-images are marked by filled circles according to the micro
magnification.  In most cases the filled circles are vanishingly
small, since most micro-images are very demagnified.  Nevertheless, it
is easy so see where the micro-images are: crossings of the contours
correspond to saddle points, while each little loop encloses a
maximum.  Each star naturally has a micro maximum at the
star position, and most have a micro saddle in the direction of the
center. This maximum/saddle micro-image configuration is dictated by
the macro shape of the arrival time, which is sloping away from the
center. Because the micro saddles have to form between the center and
micro maxima, the saddle-point contours have loops pointing
outwards. The apparent complexity of this plot should not distract
from its basic simplicity: it can be decomposed into many individual
loops, as illustrated by the accompanying sketch.
Though we have not put points at the locations of micro maxima, they do
exist inside each small loop.  Note that there are no micro minima in
this plot. Given a typical random distribution of stars it would be
rather difficult to form a local well close to the hill-top, so
micro minima are hardly ever found near macro maxima.
\citep[For an example of a microlensed macro maximum, see][]
{2007MNRAS.377..977D}

The other three panels in Figure~\ref{even-fig} show macro minima.  As
before, a star-lens forms a micro maximum, but because the whole
configuration is in a well rather than a hill, the resulting
micro saddle is in the direction away from the macro minimum. Hence
the saddle-point contours have their loops pointing towards the
macro minimum, as shown in the accompanying sketch.

Having analyzed the simplest micro-structure of a macro minimum, we
can now proceed to more involved cases. In the bottom left panel of
Figure~\ref{even-fig} we show that a single macro minimum can be split
into two with the help of a few star-lenses, which in this case happen
to form a partial ridge running vertically in the plot. In terms of
saddle-point contours, a new lemniscate, completely embedded in the
macro minimum, appears. The sketch above emphasizes that the basic
topology is that of a quad, with one or more lima\c cons surrounding
the newly formed lemniscate. 

The sequence of three panels of Figure~\ref{even-fig}, upper right,
lower left, and lower right, is primarily a sequence of nested
embedding of lemniscates within existing minima, and the accompanying
creation of additional micro minima. The lower right panel is the
latest step in that sequence, where the right micro minimum hosts a
lemniscate; the sketch above highlights the basic topology. This
sequence can be extended.  Since a minimum necessarily has $M>1$, all
micro minima contribute significant flux.  The total flux tends to
increase with the number of micro minima.  Interestingly, the
micro saddles adjacent to the micro minima also tend to be non-trivial 
in terms of flux.

In Figure~\ref{odd-fig} we show image configurations formed around
macro saddles. As before, the original macro-image would have formed
at the center of the frame. The upper left panel shows a typical
configuration with no micro minima. Star-lenses add numerous
saddle-point loops enclosing micro maxima, and adding micro saddles at
contour intersections, but these are typically faint. The sketch above it
has two configurations; the one on the left is typical; the one on the
right can replace the single point saddle-point contour
self-intersection.  Being symmetric, the configuration in the second sketch
is not stable, and hence in practice only approximations to it are seen.
A micro minimum can appear near a
macro saddle, as shown in the upper right panel. That local well is
possible to form because to the upper left and lower left of it the
macro-topology has existing ``walls'', so just one or two stars to the
right are needed to make a local well. The sketch above it also 
illustrates how a more symmetric version of a typical configuration can
create two additional micro saddles.

The lower panels of the same Figure are simple, but common
modifications; the left panel shows that similar lemniscate loops can
be added on either side of the macro saddle, and the right panel shows
that multiple, nested lemniscates can be added on the same side.
Again, we see that the micro minima are most important for the flux,
with the adjacent micro saddles also contributing significantly.

We remark that the idealized versions of our
lower left panel of Figure~\ref{even-fig} and the upper left panel
of Figure~\ref{odd-fig} are equivalent to the examples of macro minima
and macro saddles that \cite{2002ApJ...580..685S} use to explain
demagnification of saddles. 

The above discussion can be summarized as follows.
\begin{itemize}
\item The bright, and hence important images are the micro minima, and
  their adjacent micro saddles.
\item Each star-lens generates a micro maximum and a nearby
  micro saddle. Generating a micro minimum, however, requires special
  conditions: a coalition of stars in the proximity of the macro
  extremum. This is why micro minima are rare; usually just a handful
  among hundreds of micro-images.
\item Nested saddle-point structure arises naturally. One can imagine
  that if a hierarchy of lens masses is present, i.e., if the mass
  function is broad, then saddle-point contours and images of lower
  mass, smaller Einstein radius lenses can be nested within larger
  ones.  This lends another view of the bimodal lens mass model of
  \cite{2004ApJ...613...77S}.
\end{itemize}

So far we have put all the mass into stars, which average to
$\bkappa$.  However there is no loss of generality with respect to a
smoothly-distributed mass component $\kappa_c$, as the microlensing
effect of the latter can be accounted for by the simple
transformation
\begin{equation}
\bkappa \rightarrow \frac{\bkappa}{1-\kappa_c} \qquad
\bgamma \rightarrow \frac{\bgamma}{1-\kappa_c}
\end{equation}
\citep[see e.g., Equation 20 in][]{1986ApJ...301..503P}.  Such a
transformation multiplies all magnifications by the constant
$\left(1-\kappa_c\right)^2$, and is in fact equivalent
\cite[cf.][]{2000AJ....120.1654S} to the well-known mass-sheet
degeneracy in macrolensing.

Some applications of micro- and milli-lensing call for extended
perturbers.  Extended lenses can be mimicked to some extent by
softening the point lenses, which leaves the basic picture intact, and
only moves micro maxima out from under the stars.  But we do not attempt
general extended lenses in this paper.

\section{Macro magnification}

The most important observable is the macro magnification. As we saw in
the previous section, the brightest images tend to be micro minima,
which suggests that the macro magnification will depend primarily on
the number of micro mimima.  The key role of micro minima has been
noted in previous work
\citep{1992ApJ...386...30R,2002ApJ...580..685S}
but the arrival-time arguments of this paper make it evident.

To quantify our results in terms of the magnification probability
distribution we simulated $10^4$ realizations of each of the macro
minimum and macro saddle cases, with $\bar M=\pm5$ and
$\bar\kappa=0.5$ (which implies $\bgamma=0.22$ and $0.67$
respectively).  Micro-images were searched for within a square centred
on the macro-image position and containing 300 stars randomly
distributed.  In each realization, of order 300 micro saddles and up
to 4 were minima were found.

\begin{figure}
\includegraphics[width=.5\hsize]{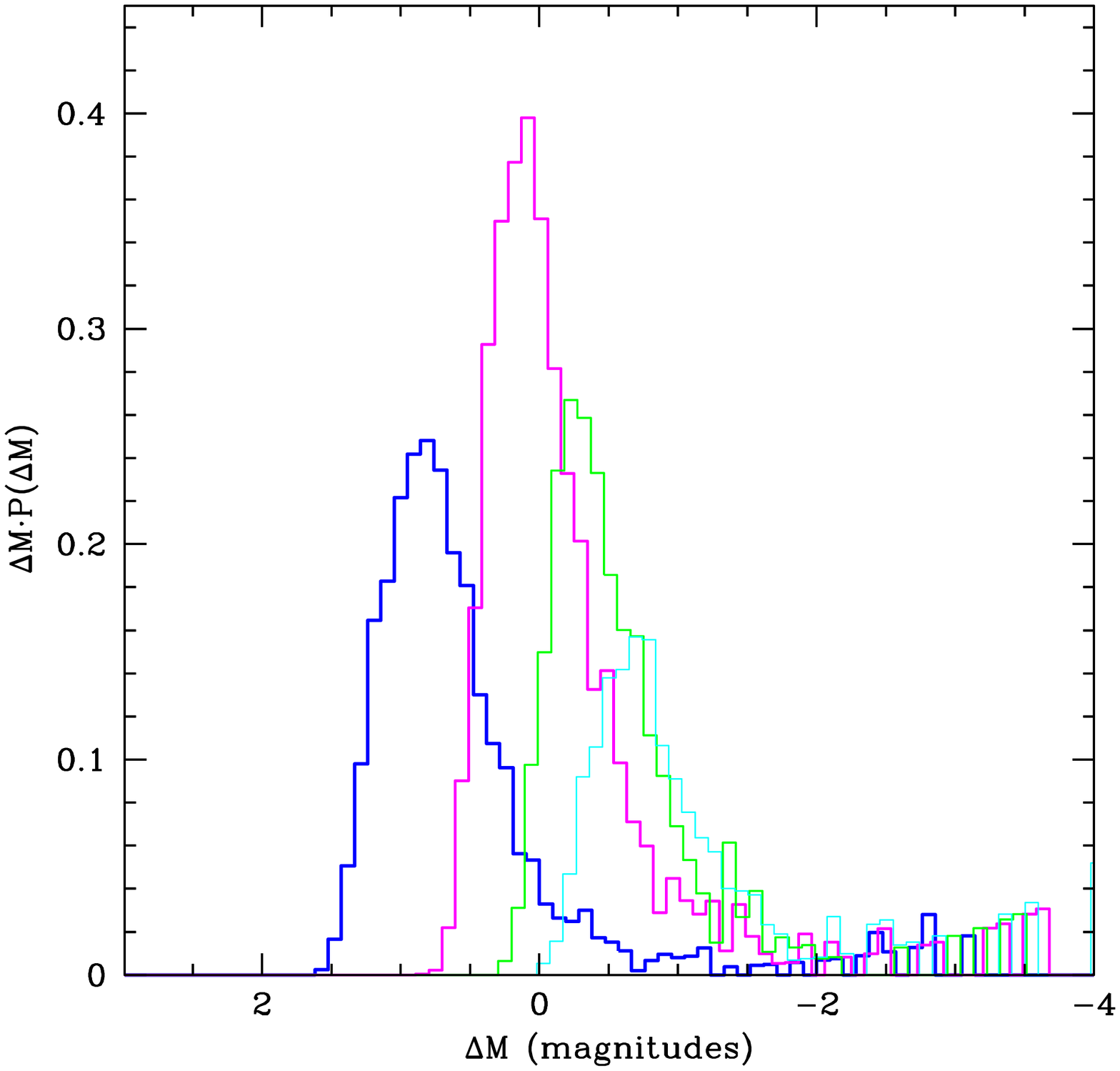}
\includegraphics[width=.5\hsize]{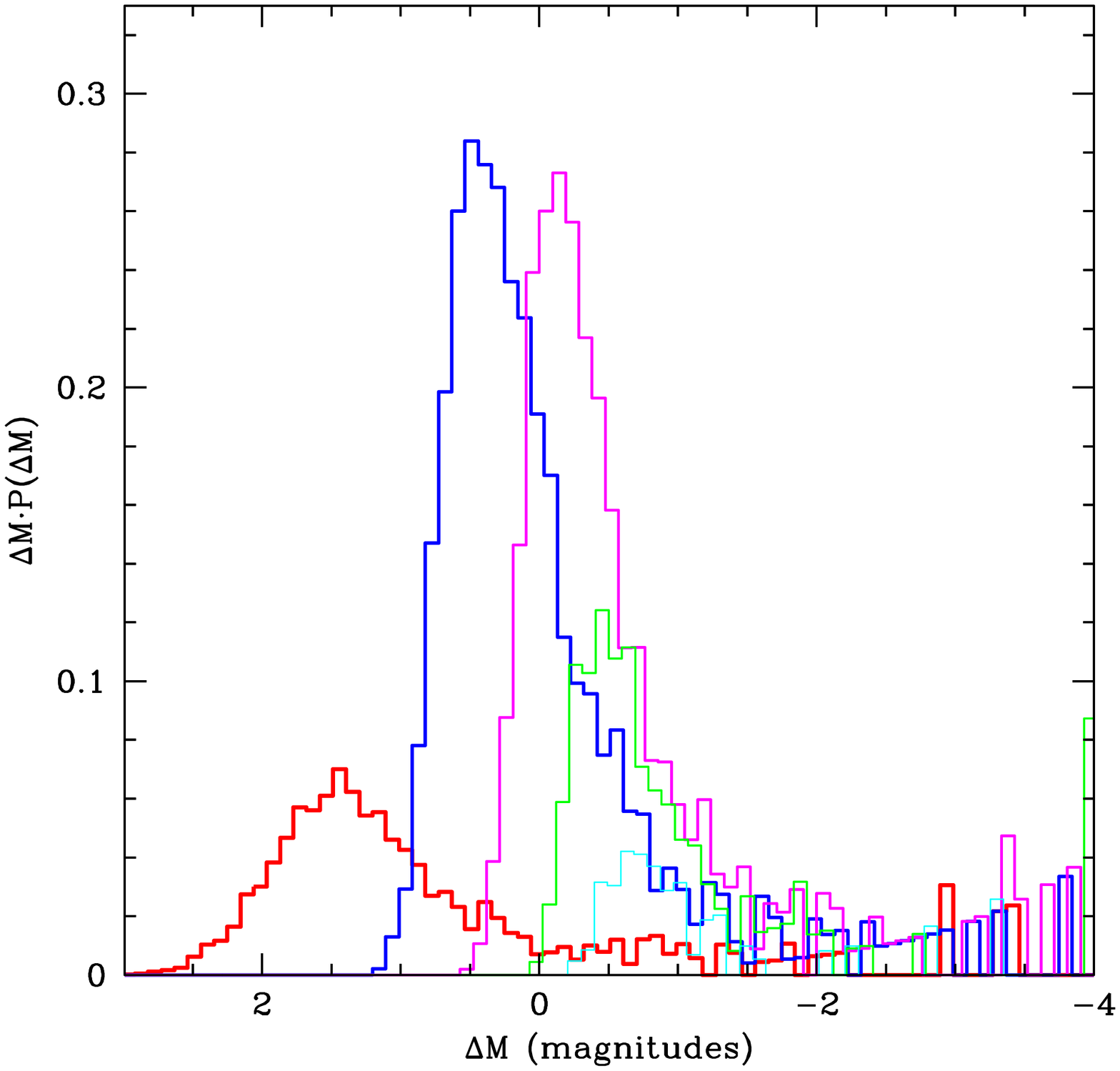}
\caption{Magnification histograms for a macro minimum (left) and
  macro saddle (right). The average magnification and surface mass density 
  are the same for both, $\bar M=\pm5$ and $\bar\kappa=0.5$, which implies 
  $\bgamma=0.22$ and $0.67$, for the two panels.
  The five distributions for the macro minimum
  show how the total magnification histogram is broken down according
  to the number of micro minima (from left to right): 1 (thickest,
  blue line), 2 (magenta), 3 (green) and 4 (thinnest, cyan line).  The
  macro saddle (right panel) is broken down in to six distributions:
  in addition to the five already mentioned, there is a still thicker
  (red) line representing macro sadxdles with 0 micro minima.  $\Delta
  M$ is total magnification (or flux) relative to the macro value.
  Note that the vertical axis is $\Delta M\,P(\Delta M)$, rather than
  $P(\Delta M)$, in order to make equal areas correspond to equal
  probabilities, since the horizontal axis is logarithmic. Vertical
  normalization is arbitrary.}
\label{histog-fig}
\end{figure}

The small number of minima suggested grouping the realizations
according to the number of minima.  Figure~\ref{histog-fig} shows
magnification distributions decomposed in this way, with the left and
right panel showing macro minima and macro saddles respectively.
Similar decompositions appear in Figure~5 of \cite{1992ApJ...386...30R}
and Figure~4 of \cite{2003ApJ...583..575G}.
It is interesting that all the sub-distributions representing cases
with at least one micro minimum have similar shapes, with a steep
dropoff on the left and a heavy tail on the right. The case of zero
micro minima (thick red histogram in the right panel) is more
symmetric.

\begin{figure}
\includegraphics[width=.5\hsize]{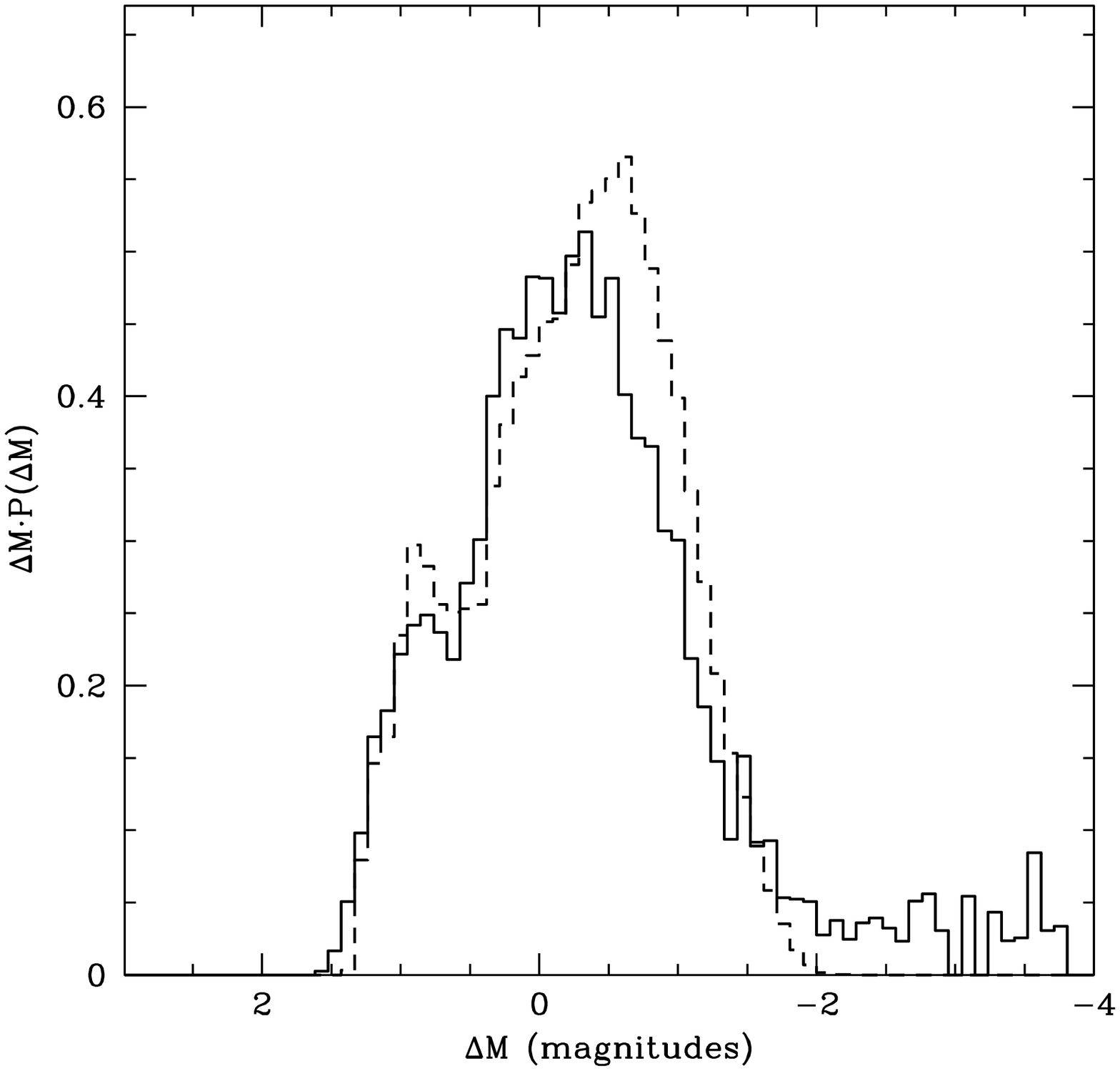}
\includegraphics[width=.5\hsize]{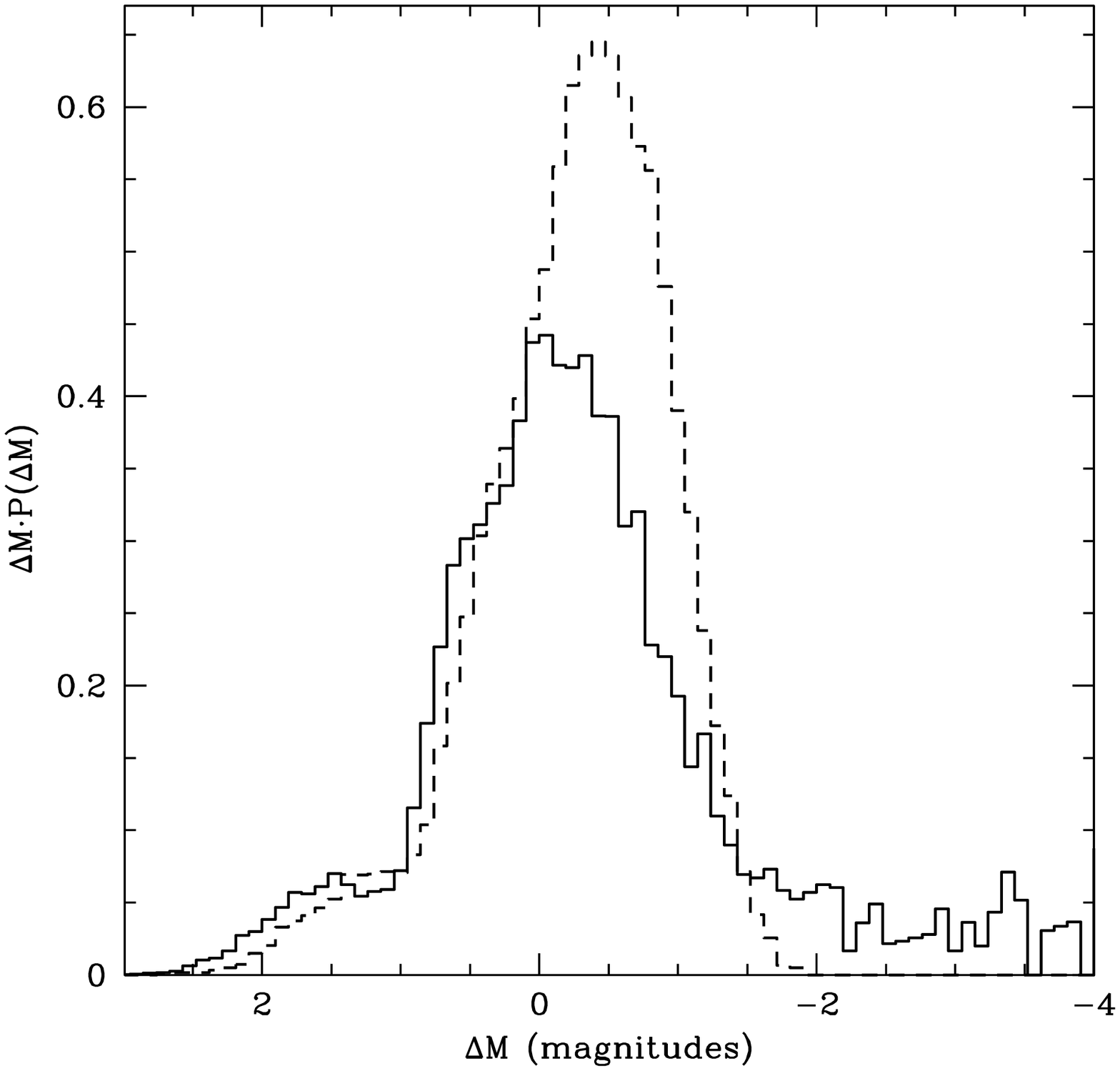}
\caption{Similar to Figure~\ref{histog-fig}, but here the solid line
  is the total of the sub-distributions presented in the corresponding
  panels of Figure~\ref{histog-fig}, while the dashed line was obtained
  from a ray tracing microlensing simulation.}
\label{histog-tot-fig}
\end{figure}

The sub-distributions overlap, but not to the degree that the
individual peaks smear out completely. The total distributions are
shown as the solid histograms in Figure~\ref{histog-tot-fig}. As a
check we compare these to the magnification distributions obtained
with ray-tracing, using a tree code.  The ray-tracing algorithm does
not find individual images, hence it does not separate out cases
according to the number of micro minima.  However, the random shear
field, Equation~\ref{microarriv}, is statistically approximated much better,
because large buffer region of rays and stars, with up to $10^5$ stars
outside the main microlensing frame is used, to eliminate edge
effects. The two distributions in each panel agree well at
macro magnification values below $\bar M$ but differ at high
magnification.  The disagreement at high magnification is expected
because the image-finding simulation assumes a point source and hence
has an very-high magnification tail not present in the ray-tracing
code, which uses a more realistic finite source. The effect of source
sizes is an interesting topic, leading for instance to chromatic
microlensing due to colour gradients in the source
\citep[see e.g.,][]{2008A&A...480..327A,2008A&A...490..933E} but we do
not investigate it in this paper.

\section{Discussion}

This paper tries to provide some new insight into micro-image
configurations and the magnification distribution arising from it.
We find that, despite the large number of micro-images and the
complexity of the pattern of images, the micro-image configurations
can be understood in terms of the arrival-time surface, as illustrated
in Figures~\ref{even-fig} and \ref{odd-fig}.  The arrival-time
contours in these figures have a very complicated structure
reminiscent of tree-rings, but can be seen to be made up much simpler
building blocks in the accompanying sketches.

The image configurations are naturally classified according to the
number of micro minima, which is almost always small ($1,2,3,4\ldots$
if the macro-image is a minimum, $0,1,2,3\ldots$ if the macro-image is
a saddle point).  Stars cannot generate micro minima on their own,
they do so in conjunction with macro shear.  The macro flux is
strongly correlated with the number of micro minima.  The well-known
broad distribution of macro fluxes can be broken down into narrow
distributions with different numbers of micro minima, as is displayed
in Figure~\ref{histog-fig}.  The main difference between the flux
distribution for macro minima and macro saddles is that the latter has
a sub-distribution with zero micro minima, which is highly
demagnified. At least in the cases where microlensing (as opposed to
lensing by extended substructure where perturbers contribute $\kappa$
as well) is important, this difference naturally explains the frequent
observational occurrence of minimum/saddle image pairs with the latter
anomalously faint.

Could the magnification distributions be understood by analytical or
partly analytical means?  This seems plausible, using the theory of
random shear introduced by \cite{1984JApA....5..235N}. The probability
distribution of the shear due to a random distribution of point masses
is \citep{1987A&A...179...80S}
\begin{equation}
P(\gamma_*) = \bkappa\,\gamma_* \left( \bkappa^2 + \gamma_*^2 \right)^{-3/2}
\label{gamma-star}
\end{equation}
The total shear $\gamma$ is the resultant of $\gamma_*$ and the macro
shear $\bgamma$ \citep[see also Equation~A17 in][]
{1984JApA....5..235N}.  The formula (\ref{gamma-star}) has led to
several interesting analytical results. Starting from
(\ref{gamma-star}), and incorporating macro shear separately,
\cite{1987ApJ...319....9S} derived the probability distribution of
high magnification events, and showed it to be independent of the
stellar mass function.  Also with the help of (\ref{gamma-star}),
\cite{1992A&A...258..591W} derived an expression for the mean number
of macro minima for the case of no macro-shear, which
\cite{2003ApJ...583..575G} later generalized to include nonzero
$\bar\gamma$.  But perhaps most relevant is the derivation by
\cite{1990ApJ...357...23L} of an analytical expression for the
magnification probability distribution, assuming a single micro-image
dominates, and there is no macro shear.  It would be very interesting
if their result could be generalized to include external shear, and
allow for several micro-images.

\section{Acknowledgements} LLRW thanks the Pauli Center for
Theoretical Studies, run jointly by the University of Z\"urich and ETH
Z\"urich, for sabbatical support.  The authors also thank the referee
for many useful comments.

\def\apj{ApJ}
\def\apjl{ApJ}
\def\apjs{ApJS}
\def\aap{A\&A}
\def\mnras{MNRAS}
\def\aj{AJ}
\def\nat{Nature}
\def\pasj{PASJ}

\bibliographystyle{mn2e}
\bibliography{ms.bbl}

\end{document}